\documentclass[conference]{IEEEtran}

\usepackage{cite}
\usepackage{amsmath,amssymb,amsfonts}
\usepackage{algorithmic}
\usepackage{graphicx}
\usepackage{textcomp}
\usepackage{xcolor}
\usepackage{booktabs} 
\usepackage{url}
\def\BibTeX{{\rm B\kern-.05em{\sc i\kern-.025em b}\kern-.08em
    T\kern-.1667em\lower.7ex\hbox{E}\kern-.125emX}}

\usepackage{draftwatermark}
\SetWatermarkText{Pre-print}
\SetWatermarkScale{1}

\begin{document}

\title{Sharing of vulnerability information among companies -- a survey of Swedish companies}

\newcommand{\linebreakand}{%
  \end{@IEEEauthorhalign}
  \hfill\mbox{}\par
  \mbox{}\hfill\begin{@IEEEauthorhalign}
}

\author{
\IEEEauthorblockN{Thomas Olsson}
\IEEEauthorblockA{\textit{Software and Systems Engineering Lab} \\
\textit{RISE Research Institutes of Sweden}\\
Lund, Sweden \\
thomas.olsson@ri.se}
\and
\IEEEauthorblockN{Martin Hell}
\IEEEauthorblockA{\textit{Dept. of Electrical and Information Technology} \\
\textit{Lund University}\\
Lund, Sweden \\
martin.hell@eit.lth.se}
\and
\IEEEauthorblockN{Martin H\"ost}
\IEEEauthorblockA{\textit{Dept. of Computer Science} \\
\textit{Lund University}\\
Lund, Sweden \\
martin.host@cs.lth.se}
\and
\IEEEauthorblockN{Ulrik Franke}
\IEEEauthorblockA{\textit{Software and Systems Engineering Lab} \\
\textit{RISE Research Institutes of Sweden}\\
Kista, Sweden \\
ulrik.franke@ri.se}
\and
\IEEEauthorblockN{Markus Borg}
\IEEEauthorblockA{\textit{Software and Systems Engineering Lab} \\
\textit{RISE Research Institutes of Sweden}\\
Lund, Sweden \\
markus.borg@ri.se}
}

\maketitle

\begin{abstract}
Software products are rarely developed from scratch and vulnerabilities in such products might reside in parts that are either open source software or provided by another organization. Hence, the total cybersecurity of a product often depends on cooperation, explicit or implicit, between several organizations.
We study the attitudes and practices of companies in software ecosystems towards sharing vulnerability information. Furthermore, we compare these practices to contemporary cybersecurity recommendations. 
This is performed through a questionnaire-based qualitative survey. The questionnaire is divided into two parts: the providers' perspective and the acquirers' perspective. 
The results show that companies are willing to share information with each other regarding vulnerabilities. Sharing is not considered to be harmful neither to the cybersecurity nor their business, even though a majority of the respondents consider vulnerability information sensitive. However, the companies, despite being open to sharing, are less inclined to proactively sharing vulnerability information.  
Furthermore, the providers do not perceive that there is a large interest in vulnerability information from their customers. Hence, the companies' overall attitude to sharing vulnerability information is passive but open. In contrast, contemporary cybersecurity guidelines recommend active disclosure and sharing among actors in an ecosystem. 
\end{abstract}

\begin{IEEEkeywords}
survey, cybersecurity, vulnerabilities
\end{IEEEkeywords}

\section{Introduction}\label{sec_intro}
Software products are rarely developed from scratch, nor by a single company or organization. Third party components can be both open source software (OSS) and purchased parts, and might depend on continuously available services from others to work as intended. The resulting product is the combined efforts from a software ecosystem~\cite{MesserschmittSzyperski2003}. Any component of the software can have vulnerabilities, whether developed internally or acquired. Vulnerabilities are ``...`flaws' or `mistakes' in computer-based systems that may be exploited to compromise the network and information security of affected systems''~\cite{ENISA}. The total cybersecurity cannot be handled by any one organization alone. Rather, it depends on the cooperation of several organizations. 

We define cybersecurity as measures taken to prevent, detect, and react to actions that may compromise confidentiality, integrity, or availability to data or devices, primarily in the context of Internet connectivity. 

There are new vulnerabilities disclosed every day. Between 2005 and 2016, the number of new vulnerabilities reported in the National Vulnerability Database (NVD)~\cite{nvd:18} ranged between 4~000-8~000 per year. In 2017 and 2018, around 14~700 and 16~500 new vulnerabilities respectively were reported. It requires both effort and knowledge to be able to evaluate these vulnerabilities, sometimes found in several different public databases, non-trivial to fuse when there is conflicting information~\cite{Johnson2018CVSS}.

The situation is further aggravated by the fact that information on a vulnerability alone is often not enough to assess whether there is an impact on the cybersecurity of a product. The configuration of the product, how the vulnerability can be exploited, the environment in which the product is running, etc., impact the damage of a vulnerability. An acquired component might use OSS with vulnerabilities. Hence, there is a need to understand not just the components developed internally but also those acquired. Even if a product might be susceptible to an exploit, it might still be more economical from a business perspective not to fix the product. For example, if the fixing and updating is expensive and the risk of the product being exploited is low. At the same time, the customers' best interest must also be considered, and it might be morally questionable to leave vulnerabilities in the software without informing them. Previous research suggests appropriate \emph{timing} is an important aspect both for vulnerability disclosure by vendors~\cite{arora2008optimal} and patching by their customers~\cite{beattie2002timing}.

In this paper, we surveyed companies' attitudes towards sharing information on their vulnerabilities within the value chain. Specifically, we studied the two perspectives of organizations \textit{providing} and organizations \textit{acquiring} software. 
\begin{enumerate}
\item[RQ1] What is the attitude towards sharing vulnerability information within the software ecosystem?
\item[RQ2] How do contemporary cybersecurity recommendations align with companies' preferences for vulnerability disclosure for (IoT) ecosystems?
\end{enumerate}

A company can either have an acquirer role or a provider role in a software ecosystem, or it has both roles. An acquirer receives software or uses a service from another company or organization, e.g. for a smart home, an acquirer acquires intelligent light switches with hardware, software, and radio. The acquirer in turn provides a product or service to others. A provider might also acquire parts of the software they provide. 
For example, the provider of intelligent light switches might develop the hardware and software for the light switches, but not the radio component. 

The rest of the paper is organized as follows: Background and related work is found in Section~\ref{sec_bgrw}. Section~\ref{sec_method} outlines the research method. The results from the survey are found in Section~\ref{sec_results} and the discussion is found in Section~\ref{sec_discussion}. The paper is concluded with Section~\ref{sec_conclusion}.

\section{Background and related work}\label{sec_bgrw}

\subsection{Background}\label{sec_bg}
The contemporary recommendations are to share information within the software ecosystem to ensure the cybersecurity of the system as a whole. However, as pointed out above, the optimal timing of disclosures is non-trivial~\cite{arora2008optimal,arora2004impact,arora2006competitive}.

The U.S. Department of Homeland Security released a document in 2016 with non-binding principles and best practices for security in connected devices~\cite{DHS:16}. They recommends that vulnerability disclosure should involve developers, manufacturers, and service providers. It should also include information regarding any vulnerabilities reported to a computer security incident response team (CSIRT). The U.S. Food and Drug Administration (FDA) has released similar recommendations~\cite{FDApre:14,FDApost:16}. The European Union Agency for Network and Information Security (ENISA) released a report in November 2017~\cite{ENISA:17}, recognizing the need for coordinated vulnerability disclosure and the importance of participating in information sharing platforms. These recommendations are also acknowledged by national bodies, e.g., the Swedish Civil Contingencies Agency (MSB) in Sweden~\cite{MSB:18}.

The Broadband Internet Technical Advisory Group (BITAG) has released recommendations stating that manufacturers should report information on software vulnerabilities that pose security or privacy threats to consumers~\cite{BITAG:16}. The IoT Security Foundation (IoTSF) state in their guidelines that an organization should have a mechanism for issuing security advisories for informing users when a problem is fixed~\cite{IoTSFVD:17,IoTSFSDBPG:18}. The Online Trust Alliance (OTA) -- an initiative within the Internet Society (ISOC) -- has released an IoT Security \& Privacy Trust Framework~\cite{OTA:17}. They recommended responsible remediate of vulnerabilities and threats. 

\subsection{Related work}\label{sec_rw}
Mufti et al. performed a systematic mapping study and a set of case studies where they develop and evaluate a ``readiness model'' for security requirements engineering~\cite{Mufti18}. The mapping study included a literature study with 104 primary studies on security requirements engineering. The resulting model describes readiness in  maturity levels from `initial' with no security requirements to `monitored' with security requirements for prevention and long-term goals. 

In our previous work, we present an interview-based qualitative survey~\cite{hostindustrial}. The purpose of that survey was to understand how security vulnerabilities, especially in third party software in IoT systems, are managed by industrial organizations. We saw that companies can be characterized according to their role in the system development value chain, from component developer to system integrator, i.e., a classification which can be used to describe sharing of information, and is similar to that in this paper. Another interview study on security and IoT is presented by Asplund et al.~\cite{Asplund16}. They investigate the degree to which security is seen as important by practitioners. They found that legacy systems enhanced by IoT solutions were often highly critical for society, which could slow down the process of transforming them into an IoT architecture, and they also found that system availability in general is more important than confidentiality of data. That is, they saw that engineers in different domains value (aspects of) security differently. 

\section{Research method}\label{sec_method}
The purpose of this qualitative survey~\cite{robson_real_2002} was to explore how companies reason about sharing information on vulnerabilities either in their own software or other companies' with whom they have a business relationship -- bilateral or other. The survey was partly descriptive and partly exploratory~\cite{easterbrook2008selecting}. It was descriptive in that the ecosystems and the companies are described. It was exploratory in that we want to build up an understanding of scenarios of how companies reason about information on vulnerabilities in general and sharing of vulnerability related information within their software ecosystem in particular.

\subsection{Study design}
In the previous work, we identified that there is a need to further understand vulnerability management within software ecosystems~\cite{hostindustrial}. In the current study, we focused on the B2B relationships among companies with either a provider or an acquirer role in the software ecosystem. 

\subsubsection{Questionnaire design}\label{sec:q_design}
We developed the first version of the questionnaire based on related work. We iterated the development of the questionnaire twice, with internal reviews among the authors in each iteration. We also asked a colleague external to the project to review the questionnaire after the second iteration. The final version of the questionnaire consisted of four parts:\footnote{The two versions of the questionnaire, including all Likert items~\cite{robson_real_2002} used can be found at \url{https://sics.box.com/s/daok9em2txe1625o2ubzl99gj5nrwora} \label{footnote_questionnaire_URL}}
\begin{enumerate}
    \item Five context questions to characterize the companies.
    \item Five groups of Likert items\footnote{A Likert item is one statement that the respondent should judge whether they agree or disagree to it.} on vulnerability management and related practices for providers.
    \item Five groups of Likert item on vulnerability management and related practices for acquirers. 
    \item Four general fix-answers questions on the companies' handling of cybersecurity. 
\end{enumerate}

The attitude towards being more open to share information regarding vulnerabilities was investigated. Hence, a high sum total score on the Likert items should indicate a willingness to share information with others and a low score that the companies are less inclined to do so. The two parts for providers and acquirers are mirrored as far as possible. If there is a Likert item for the provider part, it is usually also found in the acquirer part. This means we can compare the perspectives. Companies can be both providers and acquirers at the same time, though this is not always the case. The five groups of Likert items make up the Likert scales found in Table~\ref{Tab:Likertgroup}.

\begin{table}[]
\caption{The groups of Likert items and their corresponding identifiers in the provider and acquirer part\textsuperscript{\ref{footnote_questionnaire_URL}} -- see also Figure~\ref{fig:Likert_scale} and Figure~\ref{fig:Large-NonLarge}. }
\label{Tab:Likertgroup}
\begin{tabular}{r|cc}
                                                                                                           & \multicolumn{2}{c}{Identifying number} \\
Likert item groups                                                                                         & Provider       & Acquirer       \\ \hline
Practices and prevalence of OSS                                                                           & 6              & 12            \\
Description of the business ecosystem                                                                     & 7              & 13            \\
Information sharing in the value chain                                                                    & 8              & 14            \\
\begin{tabular}[c]{@{}r@{}}Effects of information sharing \\ in the value chain\end{tabular}           & 9              & 15            \\
\begin{tabular}[c]{@{}r@{}}Practices related to software \\ update and release management.\end{tabular} & 10             & 16           
\end{tabular}
\end{table}

We decided to have five fixed-alternative expressions, ranging from ``strongly disagree'' to ``strongly agree''~\cite{robson_real_2002}. We also decided to include a "do not know/not applicable" alternative as it is plausible that not all respondents can answer all items in the Likert questions~\cite{fowler_survey_2013}. Lastly, three items in each of the two parts were deliberately phrased negatively in the sense that a disagreement should be treated as an agreement and vice verse. 

\subsubsection{Execution}
In the first data collection round, we asked members of a cybersecurity project to fill in the survey. In this step, we selected the companies purposely~\cite{robson_real_2002} as we knew their interest and maturity in cybersecurity. The respondents were security experts from the large and medium sized companies and managers from the small companies. 

Based on the first round, we made some adjustments to the questionnaire. Specifically, we added one Likert item to two groups for the provider part (6f and 9c) and to the two corresponding groups for the acquirer part (12f and 15c). In addition, we removed one Likert item from one question for the provider part\textsuperscript{\ref{footnote_questionnaire_URL}} (13d in the first version of the questionnaire). As the changes are minor, we see no risk in using those answers along with the rest of the answers. 

For the second round, we first used a systematic sample with large companies in Sweden\footnote{We selected the companies registered at the Swedish stock exchange on the large capital list. At the time of the study, it was 94 companies. However, we excluded pure investment companies. Therefore our sample frame was 84 companies.} as our sample frame~\cite{fowler_survey_2013}. We approached the companies through their web page and contact information which could be found there. We asked to get in touch with CTOs, business managers, etc., within the companies. We employed a person on an hourly basis to elicit the answers from the companies. They contacted companies both via e-mail and by phone. Once a contact was established -- beyond the generic switchboard -- we reminded them every second week and we sent them up to 5 reminders. We got answers from managers either specifically for security or for IT. 

In the third round, we used the same questionnaire as in the second round. To complement the answers gathered in the first and second round, we utilized our professional networks to elicit additional answers, i.e., convenience sampling~\cite{fowler_survey_2013}. The respondents were more mixed but overall individuals with good insight into the software development of their companies but not necessarily cybersecurity. 

\section{Results and Analysis}\label{sec_results}

\subsection{Sample companies}\label{sec_companies}
In total, we got answers from 17 companies of different sizes and various domains, see Table~\ref{tab:Companies}. 5 companies answered only the provider part, 4 companies answered only the acquirer part and 8 companies answered both parts. 

\begin{table}[htb]
\caption{List of companies and their size, in terms of number of employees.}
\label{tab:Companies}
\begin{tabular}{@{}r|cccc|cc@{}}
Round  & \begin{tabular}[c]{@{}c@{}}Large \\ $>250$ \end{tabular}  & \begin{tabular}[c]{@{}c@{}}Medium \\ $<250$ \end{tabular}  & \begin{tabular}[c]{@{}c@{}}Small \\ $<50$ \end{tabular} & \begin{tabular}[c]{@{}c@{}}Micro \\ $<10$\end{tabular} & Total  \\ \midrule
First  & 1\textsuperscript{\ref{footnote:Twice}}     & 1      & 2     &  & 4\textsuperscript{\ref{footnote:Twice}}                     \\
Second & 3     &        &       &  & 3                                                        \\
Third  & 7     &        &       & 3 & 10                                                        \\ \midrule
Total  & 10\footnote{Test}    & 1      & 2     & 3  & 17\textsuperscript{\ref{footnote:Twice}}                                                    
\end{tabular}
\end{table}
\footnotetext{The large company in round one also provided an answer in round two. When the answer is removed, we have answers from 9 large companies and in total 16 companies, cf. Section~\ref{sec_companies}, second to last paragraph.\label{footnote:Twice}}

In the first round, there were 4 companies . We got answers for all of them, as expected since we knew they had an interest in the subject. 

In the second round, 84 companies from our sample frame contacted. 18 companies could not find an appropriate person who was willing to answer the questionnaire. 52 companies never replied at all. Our initial email bounced for 9 companies and we could not find other ways to contact them. Three companies explicitly said they would not answer the survey as they neither provide nor acquire software-intense products relevant for our study. We got answers from 3 companies. 

In the third round, we used convenience sampling. In total, we contacted 22 companies. This resulted in 10 answers.

The large company from round one also provided an answer in round two. However, the answer was provided by different individuals, though from the same business unit. The answers are very similar. Therefore, we decided to keep the answer from round two, as round two is from a methodology perspective more rigorous. 

All of the companies are active in Sweden, operating on an international market. All of the large companies have development units both within and outside Sweden whereas the non-large companies have their development work in Sweden. 

\subsection{Final Likert scale}\label{sec:finalscale}
In order to select the items of the Likert scale to use in the analysis, we perform an item analysis to identify which of the items that have discriminative power~\cite{robson_real_2002}. (As three items in each part of the survey are phrased negatively, we first inverted those answers before performing the item analysis.) Those items with the largest discriminative power are those that best answer the overall question of the attitude of sharing vulnerability information. The following items are removed: 
\begin{itemize}
    \item The items related to the business ecosystem both for the provider (group 7) and the acquirer part (group 13) have a low discriminative power. 
    \item One item from the OSS group in the provider part has no counterpart in the acquirer part and low discriminative power (item 6e). 
    \item For release management and software updates, one item (10b) in the provider part has no corresponding item in the acquirer part and a low discriminative power. 
    \item Two items (10c and 10e and the corresponding 16c and 16d) have a low discriminative power in both parts.
\end{itemize}

One item in the acquirer part (12e) has no corresponding item in the provider part but a high discriminative power. That item is retained in the final scale. The final Likert scale is found in Figure~\ref{fig:Likert_scale}. 

\begin{figure*}[tb]
\includegraphics[width=1\textwidth]{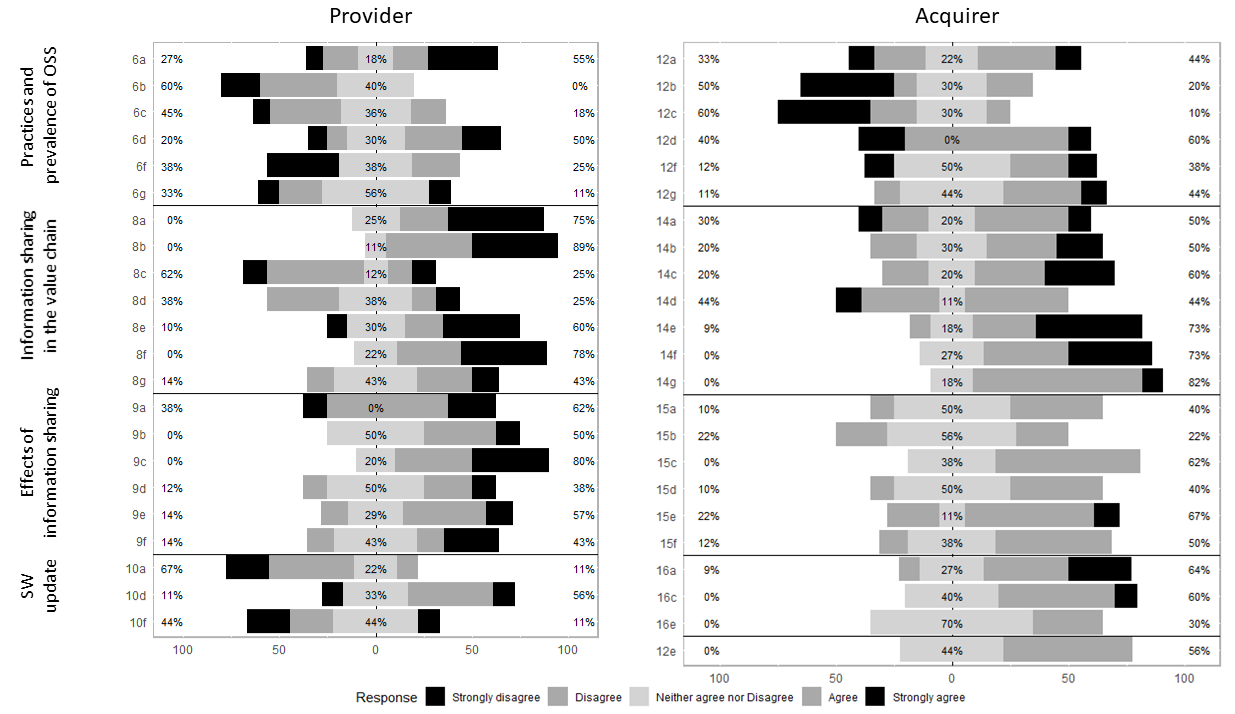} 
 \caption{Providers and acquirers, stacked bar charts representing the frequency of answers per Likert item (i.e. each item sums up to 100\%). Note that the answers are inverted for 6 items as the questions are phrased negatively, see Section~\ref{sec:finalscale}. Also note that item 12e does not have a corresponding item in the provider part.}
 \label{fig:Likert_scale}
\end{figure*}

\begin{figure*}[tb]
\includegraphics[width=0.9\textwidth]{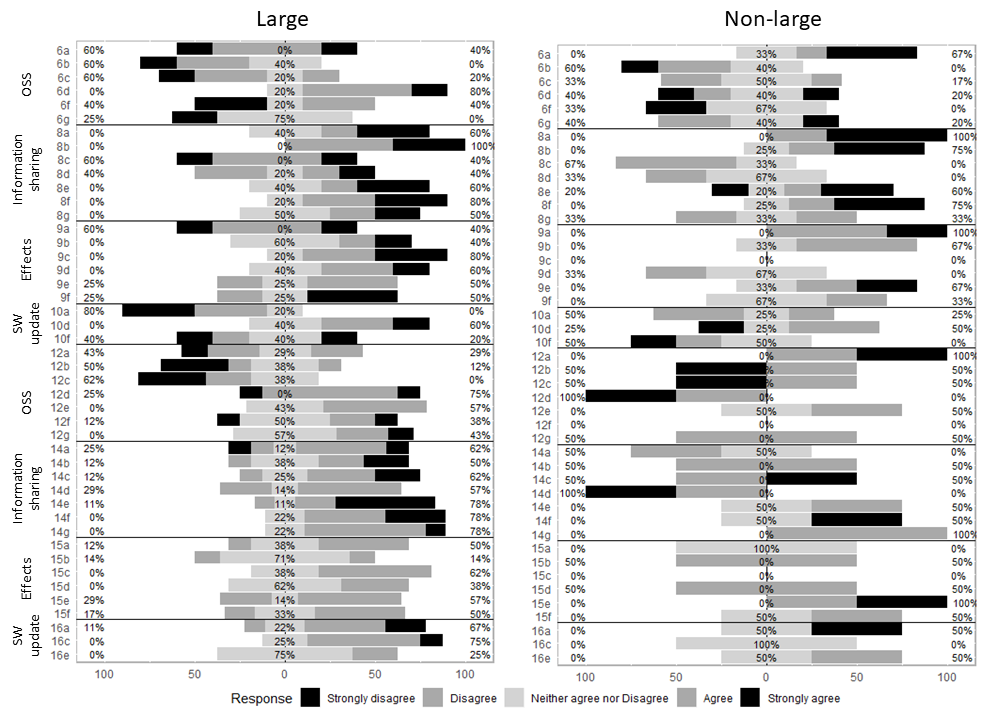} 
 \caption{Large and non-large companies, stacked bar charts representing the frequency of answers per Likert item (i.e. each item sums up to 100\%). (Non-large are the medium, small and micro companies.) Note that the answers are inverted for 6 items as the questions are phrased negatively, see Section~\ref{sec:finalscale}.}
 \label{fig:Large-NonLarge}
\end{figure*}

\subsection{Analysis}
\subsubsection{Practices and prevalence of OSS}
OSS is a significant part of the products for many companies in the study (cf. 6a and 12a in Figure~\ref{fig:Likert_scale}). However, it is a larger part in the non-large (medium, small, and micro) companies than in large companies, see Figure~\ref{fig:Large-NonLarge}. Hence, smaller companies seem to utilize OSS more than larger companies in the survey. However, the companies are less inclined to actively contribute to OSS projects (6b, 6c, 12b and 12c). Non-large companies seem slightly more inclined to contribute than large companies. The difference is smaller than the difference on prevalence on OSS. 

Another similarity between providers and acquirers is that non-large companies seldom have dedicated individuals who monitor forums on cybersecurity for OSS, e.g., NVD and exploit-DB (6d and 12d in Figure~\ref{fig:Likert_scale}). It seems like larger companies in the survey are more inclined to have dedicated individuals monitoring vulnerabilities databases, independent of role (6d and 12d in Figure~\ref{fig:Large-NonLarge}). 

Acquirers use more third-party software or service to keep track of vulnerabilities in OSS used (6f and 12f) than providers. Also, the acquirers seem to demand that providers keep their OSS component updated with the latest version -- more than the providers are inclined to provide it (6g and 12g). There is no difference between large companies and non-large companies in the survey.

\subsubsection{Information sharing in the value chain}
The providers answer more positive to sharing vulnerability information with their customer (8a). The acquirers indicate that they do not get security information from their providers (14a) -- with a similar pattern for critical vulnerabilities (8c and 14c in Figure~\ref{fig:Likert_scale}). Non-large acquirers answered, in the median, around ``Neither agree nor disagree", see Figure~\ref{fig:Large-NonLarge}. 

Both acquirers and providers are inclined to ask for information on vulnerabilities. The acquirers seem less inclined to ask the providers for information on security whereas the providers comply with requests from acquirers on security (8b and 14b in Figure~\ref{fig:Likert_scale}) -- no difference whether a large or non-large company. This is in line with that the providers perceive an interest in cybersecurity from their customers (8e and 14e in Figure~\ref{fig:Likert_scale}) -- which acquirers also answer. Both providers and acquirers are aligned in the perception that cybersecurity information is sensitive and that vulnerabilities addressed should be communicated in, e.g., release notes (8f, 8g, 14f and 14g in Figure~\ref{fig:Likert_scale}) -- no difference between large and non-large companies.

\subsubsection{Effects of information sharing in the value chain}
The items related to the information sharing practices -- groups 8 and 14 -- are contrasted by the items related to the effect of information sharing in the value chain -- groups 9 and 15. For the latter, the attitude in general is more positive to sharing than compared to the actual practices from the former. Most answers are either neutral or positive, cf. Figure~\ref{fig:Likert_scale} (note that 9a and 15a are negatively posed and should be interpreted inversely). The exception is the item asking whether sharing cybersecurity information is harmful (9a), where the attitude from the providers is more balanced. Furthermore, non-large providers are less inclined to consider vulnerability information to be harmful than larger companies, see 9a in Figure~\ref{fig:Large-NonLarge}. The profile is similar between providers and acquirers, albeit the latter is less extreme in their answers. 

\subsubsection{Practices related to software update and release management}
Both acquirers and providers indicate in their answers that updating software is costly (10d and 16c). However, the acquirers still have an ambition to keep the software up to date (16a). Providers, when asked if they update their software frequently (daily or weekly) indicated that they do not (10a). However, as that item is expressed using the extreme value of daily or weekly, the results should be interpreted with care. Lastly in the group on release management, the acquirers are more positive to pay for updates than the providers perceive (10f and 16e). The answers from large and non-large companies are, in the median, similar. 

\subsubsection{General questions on cybersecurity}
41\% (7) answered that they are using a third-party service to handle cybersecurity and 23\% (4) answered that they have not considered it (question 17). In terms of cybersecurity competence, 9 answered that they have competence within the company to handle cybersecurity and 8 that they do not (question 18). Cyber insurance is only used by 2 of the companies in the study and 10 of the companies have not even considered it (question 19). This is not surprising, as it is known that cyber insurance uptake in Sweden is still low \cite{franke2017cyberinsurance}. Lastly, 9 of the companies find cybersecurity to be very important whereas 8 do not prioritize it (question 20). There is no correlation among these questions (17-20). 

\section{Discussion}\label{sec_discussion}
Half of the companies in the survey answered that they only have a basic understanding of cybersecurity, which seems to be uncorrelated to how important cybersecurity is perceived. The sample is likely skewed, however, as it is plausible that people are more inclined to participate in the survey if they have an interest in cybersecurity. This is further reinforced by our difficulties to elicit answers, especially in round two. Despite a substantial investment with an hourly employed person to elicit answers, we only got a 3.6\% response rate in this round, even though several of the companies indicate that they find cybersecurity important. We see three possible explanations for the difficulties to elicit answers, in addition to survey-fatigue and scarcity of time that affects all surveys: 1) even though the spontaneous answer is that cybersecurity is important, in reality respondents are not prepared to actually invest in it, 2) the topic of the survey is advanced and many organizations simply lack the experience and competence to answer the questionnaire, and 3) the topic is considered too sensitive to answer. The first two explanations might indicate a need to raise the cybersecurity awareness and competence in industry.

The prevalence and approach to OSS is similar between providers and acquirers. However, as the acquirers are not updating the software with the latest OSS version themselves, they expect the providers to take the cost. Also, as the acquirers are likely less capable of handling vulnerabilities on the OSS components, they are more inclined to use third-party tools or services to monitor the security. Hence, the awareness of the importance of having a proactive attitude to updated software seems to falter. 

\subsection{RQ1 What is the attitude towards sharing vulnerability information within the software ecosystem?}
In general, companies seem reactive rather than proactive when it comes to information sharing of vulnerabilities. There is a willingness to share information, at least with business partners (8a, 8b, 14a, and 14b), however, it does not seem to be proactive and planned upfront (8c, 8d, 14d). At the same time, the attitude to sharing is positive as seen in groups 9 and 15. This indicates that the perception of wanted position and the actual reality in the companies is not aligned. This is similar to the handling of OSS. The companies commonly use OSS (items 6a and 12a), however, are less inclined to contribute back or actively participating in the community (items 6b, 6c, 12b and 12c). This indicates that the companies do not have an elaborate OSS or software ecosystem strategy, whether large or non-large. Rather, it seems as if they want to get access to assets but do not see the benefit from giving away assets without monetary compensation. This indicates that the companies in the survey have not yet seen the benefit of sharing, indicating an immaturity. Interestingly, this is true even for non-large companies, which we expected to be more active and involved in OSS communities. 

Both providers and acquirers consider vulnerability information to be sensitive (items 8f and 14f). When asked whether they do not share cybersecurity information because sharing it can be harmful (items 9a and 15a), the attitude seems contradictory to the previous question. 
We interpret this as an indication that respondents are unsure how sensitive it is to share vulnerability information. There might also be an influence from the wording -- and therefore the interpretation -- of the questions on the answers. Interestingly, the attitude to being transparent and open with vulnerabilities is overall positive (9b and 15b). This can be contrasted to the perception that the counterparts in the software ecosystem are negative to disclosing vulnerability information (9f and 15f respectively). We speculate that this is due to self-serving bias, where oneself is considered overly mature and counterparts overly immature. Overall, in relation to the research question, we hypothesize that there is a lack of maturity and therefore practices for how to handle vulnerability disclosure. 

In relation to RQ1, we interpret the answers that overall the companies do not have explicit nor elaborate practices regarding sharing of vulnerability information. Furthermore, the companies seem, in general, to consider vulnerability information to be sensitive, though this is not a well-informed opinion. 

\subsection{RQ2 How do contemporary cybersecurity recommendations align with companies' preference for vulnerability disclosure for (IoT) ecosystems?}

As outlined in the background section (Section~\ref{sec_bg}), government agencies and industry interest groups in general promote disclosure and sharing of vulnerability information in the software ecosystem. Both American and European government agencies recommend (proactive) and even coordinated disclosure~\cite{DHS:16,ENISA:17}. The idea is that if the actors in a software ecosystem cooperate, cybersecurity will be improved. The BITAG and ISOC organizations go even further and indicate that threats should be reported to the consumers as well~\cite{BITAG:16,OTA:17}. In our survey, providers tend to be less proactive in providing information but do provide it if requested. On the other hand, acquirers are not that interested in cybersecurity. Also, vulnerability information is considered sensitive -- at least by providers. Hence, there seems to be a tendency to keep information confidential as a way to achieve ``security through obscurity". Therefore, even the respondents who rate themselves as competent in the area of cybersecurity seem to be misaligned with contemporary recommendations. We speculate that this is a combination of lack of competence in cybersecurity as well as a lack of understanding how to communicate around vulnerabilities as it can be seen as negative from the market~\cite{BHARADWAJ200966} and customer perspectives. Indeed, vulnerability disclosure could be a tragedy of the commons, where the recommendations are correct that overall cybersecurity would increase if there was more information sharing, but individual companies still could be worse off implementing such practices, at least in the short run. However, it could also be that information sharing is good for individual companies even in the short run, but that they fail to appreciate this, e.g., because the investments needed to become more mature are more tangible than the benefits entailed. These lines of reasoning imply that it is not just about increasing the cybersecurity knowledge in the software development parts of a company but a more general problem that requires a broader approach outside the technical roles.

There are also recommendations, e.g., from ISOC~\cite{OTA:17}, on the software update release process. Acquirers seem quite willing to update often and even pay for updates. The providers seem less inclined to providing updates with the same frequency. Most of the cost might be on the provider, explaining this. It can also be, however, that providers have a deeper and more nuanced insight into the actual need for updates from a technical perspective, whereas acquirers simply always want ``latest and greatest''. This discrepancy indicates that there is a mismatch in the software ecosystem in understanding each other in terms of incentives and willingness to pay for updated software. For cybersecurity, this might imply that the lack of communication and mutual understanding of the actual need for addressing vulnerabilities lead to unnecessary cost and efforts. At the same time, the acquirers are somewhat inclined to trust their providers (12e in Figure~\ref{fig:Likert_scale}). This implies that there is a lack of practices and an underlying assumption that the counterparts in the software ecosystem can be trusted.

\subsection{Threats to validity and reliability}
Here the validity of the conducted research is discussed based on commonly considered validity threats, e.g., as listed in \cite{KitchenhamPfleeger}.

Content validity concerns how appropriate the contents of the survey questions are to the respondents. The questions asked concern cybersecurity, which requires some level of competence to answer. Because of that, measures were taken not only to formulate the questions in a clear and understandable way with terms used in industry, but also to find the right persons in the organizations to answer the questions. In the first round, questions were answered by respondents participating in a research project on cybersecurity, which means that uncertainties in the questions could be solved. In the later rounds initial contacts were often taken with subjects in management positions, but they were asked to get support by experts in their organizations. During the research it was found that this type of questions can be difficult to answer, at least when seeking an answer for the entire company. This emphasizes the importance of formulating the questions as clearly as possible, and spending effort on finding the right respondents. We believe the measures we have taken appropriately address the threats to content validity. 

Construct validity concern the degree to which the constructs investigated are actually measured by the questions. In this type of study this is a threat since there is a risk that people do not interpret the questions in the way intended by the researchers, especially when they cover aspects such as cybersecurity that requires some specific competence to understand. Well formulated questions increase both the number of possible respondents and the likelihood that they understand the content of the questions. The questions were developed in iterations, which we believe resulted in questions with lowered risk of misinterpretations. Also, in the first round they were answered by respondents in a research project on cybersecurity, which we believe made these respondents more motivated to understand the questions and to give feedback than would be the case if we had started with respondents from the later rounds. 

Reliability concerns the degree to which a respondent would hypothetically give the same answers if they answered the same questions again under the same conditions, or if two persons would give the same answers, e.g., measured by a measure of rater agreement. We have not measured the agreement between respondents since the sample represents different views. If the sample would be significantly larger it may have been possible to study groups of respondents, but with this sample size it is not realistic. 

The validity of a survey is, of course, affected by the possibility to generalize from the sample to the entire sample frame. We cannot argue to have a statistically rigorous sampling approach (cf. Section~\ref{sec_method}). However, our main aim is not to quantitatively survey and make generalization statements of a population. Rather, our aim is to understand the diversity and different approaches, i.e. the same reason we do not investigate rater agreement when we analyze reliability. Hence, we are not attempting to generalize the findings to an entire population. However, at the same time, the threats to generalization should not be exaggerated~\cite{flyvbjerg2006a}. As we have covered different company sizes, several domains, both technical and non-technical ones, and many locations in Sweden, we believe the findings to be, at least to some extent, relevant for providers and acquirers of software-intense products. We should have some coverage of relevant phenomena in that domain even though we cannot make statements of the statistical significance. 

For the larger companies there may be several more or less independent parts of the companies. Hence, for those cases, the answers are more from that part of the company rather than for the entire company. Again, this of course is a threat to generalization. At the same time, even though results are not statistically significant, we do believe the findings in general are relevant for the entire population.

\section{Conclusion}\label{sec_conclusion}
The attitude towards open communities and sharing of information is, in general, positive. However, when asked specific questions, respondents' attitude seems to be that it is OK to use information and source code from others but the companies are less positive to actually contributing themselves, whether to OSS development or to vulnerability information disclosure. In combination with what seems to be a lack of established practices and procedures on vulnerability disclosure and communication, we identify a need to further improve the competences required for this. Furthermore, we call for improved processes and decision support within product management and market communication, etc., as vulnerability handling is not isolated to purely technical issues. Lastly, there might be a further need to adapt and establish recommendations from government agencies and industry interest groups.

The practical implication of this study is that there is a need for companies to increase their knowledge of vulnerability management in general and specifically understanding their own technical environment to be able to make informed decisions on how to analyze and share vulnerability information. There also seems to be a role for a trusted third party to facilitate the sharing of vulnerability information. However, we believe there is a need to better understand the vulnerability sharing in order to provide validated guidelines. Future research should address business models related to vulnerability information sharing. Furthermore, we believe there is a need to better understand how individual vulnerabilities in individual parts of a larger software system configuration impacts the overall cybersecurity. Lastly, there is a need to better understand how to upfront design complex systems made up of components from several companies and OSS communities to allow for analyzability of the resulting cybersecurity. 

\section*{Acknowledgements}
We would like to thank all the participating companies and Vinnova (grant 2016-00603 and grant 2018-03965) for funding our research. U.~Franke was partially supported by the Swedish Civil Contingencies Agency, MSB (agreement no. 2015-6986). We would also like to thank Daniel Wisenhoff for his work in collecting answers. 

\bibliographystyle{IEEEtran}
\bibliography{references}  

\end{document}